\documentclass[pra,twocolumn,a4paper]{revtex4}
\usepackage{amsmath,amssymb,graphicx,color,amscd}

\newcommand{\traceSubInline}[2]{\textrm{Tr}_{#1} \! (#2)}

\newcommand{\integ}[1]{\ensuremath{\int \!\! \mathrm{d}#1 \,}}
\newcommand{\integlim}[3]{\ensuremath{\int_{#1}^{#2} \!\!\! \mathrm{d}#3 \,}}

\newcommand{\deriv}[2]{\frac{\textrm{d} #1}{\textrm{d} #2}}

\newcommand{\mean}[1]{\ensuremath{\left\langle #1 \right\rangle}}

\newcommand{\prob}[1]{\textrm{Pr} \! \left(#1\right)}

\newcommand{\nbar}{\ensuremath{\bar{n}}}

\newcommand{\xzp}{\ensuremath{x_0}}
\newcommand{\xref}{\ensuremath{x_\textrm{lin}}}
\newcommand{\xdis}{\ensuremath{x_\textrm{sq}}}

\newcommand{\omegam}{\ensuremath{\omega_M}}
\newcommand{\omegaL}{\ensuremath{\omega_L}}

\newcommand{\gref}{\ensuremath{g_{\textrm{lin}}}}
\newcommand{\gdis}{\ensuremath{g_{\textrm{sq}}}}

\newcommand{\chiX}{\ensuremath{\chi_{\scriptscriptstyle{\! X}}}}
\newcommand{\chiDis}{\ensuremath{\chi_{\textrm{sq}}}}

\newcommand{\OmegaRefX}{\ensuremath{\Omega_{\textrm{lin}}}}
\newcommand{\OmegaDis}{\ensuremath{\Omega_{\textrm{sq}}}}

\newcommand{\Fref}{\ensuremath{\mathcal{F}_\textrm{lin}}}
\newcommand{\Fdis}{\ensuremath{\mathcal{F}_\textrm{sq}}}

\newcommand{\Href}{\ensuremath{H_{\textrm{lin}}}}
\newcommand{\Hdis}{\ensuremath{H_{\textrm{sq}}}}

\newcommand{\UpsXL}{\ensuremath{\Upsilon_{\! X}^{\phantom{\dagger}}}}
\newcommand{\UpsdXL}{\ensuremath{\Upsilon_{\! X}^\dagger}}

\newcommand{\UpsDis}{\ensuremath{\Upsilon_\textrm{\!\! sq}^{\phantom{\dagger}}}}

\newcommand{\ad}{\ensuremath{a^\dagger}}
\newcommand{\bd}{\ensuremath{b^\dagger}}

\newcommand{\alphai}{\ensuremath{\alpha_{\textrm{in}}}}

\newcommand{\ai}{\ensuremath{a^{\phantom{\dagger}}_{\textrm{in}}}}

\renewcommand{\ao}{\ensuremath{a^{\phantom{\dagger}}_{\textrm{out}}}}
\newcommand{\aod}{\ensuremath{a^\dagger_{\textrm{out}}}}

\newcommand{\alphalo}{\alpha_{\textrm{LO}}}

\newcommand{\XM}{\ensuremath{X_M}}
\newcommand{\PM}{\ensuremath{P_M}}

\newcommand{\XMin}{\ensuremath{X_M^{\textrm{in}}}}
\newcommand{\PMin}{\ensuremath{P_M^{\textrm{in}}}}

\newcommand{\XMout}{\ensuremath{X_M^{\textrm{out}}}}
\newcommand{\PMout}{\ensuremath{P_M^{\textrm{out}}}}

\newcommand{\XMval}{\ensuremath{X_M}}

\newcommand{\XLout}{\ensuremath{X_L^{\textrm{out}}}}

\newcommand{\homX}{\ensuremath{Q_{\!X}}}
\newcommand{\homXZero}{\ensuremath{Q_{\!X}^{(0)}}}
\newcommand{\homP}{\ensuremath{Q_{\!P}}}
\newcommand{\delhomX}{\ensuremath{\Delta Q_{\!X}}}

\newcommand{\rhoM}{\ensuremath{\rho_M}}
\newcommand{\rhoMin}{\ensuremath{\rho_M^{\textrm{in}}}}
\newcommand{\rhoMout}{\ensuremath{\rho_M^{\textrm{out}}}}

\newcommand{\etal}{\emph{et al}.}
\newcommand{\PRL}[4]{``#4," Phys. Rev. Lett.~\textbf{#1}, #2 (#3)}
\newcommand{\PRA}[4]{``#4," Phys. Rev. A~\textbf{#1}, #2 (#3)}
\newcommand{\PRAR}[4]{``#4," Phys. Rev. A~\textbf{#1}, #2(R) (#3)} 

\newcommand{\Nature}[4]{``#4," Nature (London)~\textbf{#1}, #2 (#3)}
\newcommand{\NatPhys}[4]{``#4," Nature Physics~\textbf{#1}, #2 (#3)}
\newcommand{\NatPhot}[4]{``#4," Nature Photonics~\textbf{#1}, #2 (#3)}
\newcommand{\Science}[4]{``#4," Science~\textbf{#1}, #2 (#3)}
\newcommand{\Physics}[4]{``#4," Physics~\textbf{#1}, #2 (#3)}
\newcommand{\APL}[4]{``#4," Appl. Phys. Lett.~\textbf{#1}, #2 (#3)}
\newcommand{\NJP}[4]{``#4," New J. Phys.~\textbf{#1}, #2 (#3)}

\begin{document}

\title{Selective linear or quadratic optomechanical coupling via measurement}

\author{M. R. Vanner}
\affiliation{
Vienna Center for Quantum Science and Technology (VCQ),
\mbox{Faculty of Physics, University of Vienna, Boltzmanngasse 5, Vienna A-1090, Austria}}

\date{\today} 

\begin{abstract}
The ability to engineer both linear and non-linear coupling with a mechanical resonator is an important goal for the preparation and investigation of macroscopic mechanical quantum behavior. In this work, a measurement based scheme is presented where linear or square mechanical displacement coupling can be achieved using the optomechanical interaction linearly proportional to the mechanical position. The resulting square displacement measurement strength is compared to that attainable in the dispersive case using the direct interaction to the mechanical displacement squared. An experimental protocol and parameter set are discussed for the generation and observation of non-Gaussian states of motion of the mechanical element.
\end{abstract}

\maketitle

\section{Introduction}

\begin{figure}[t h]
\includegraphics[width=0.9\hsize]{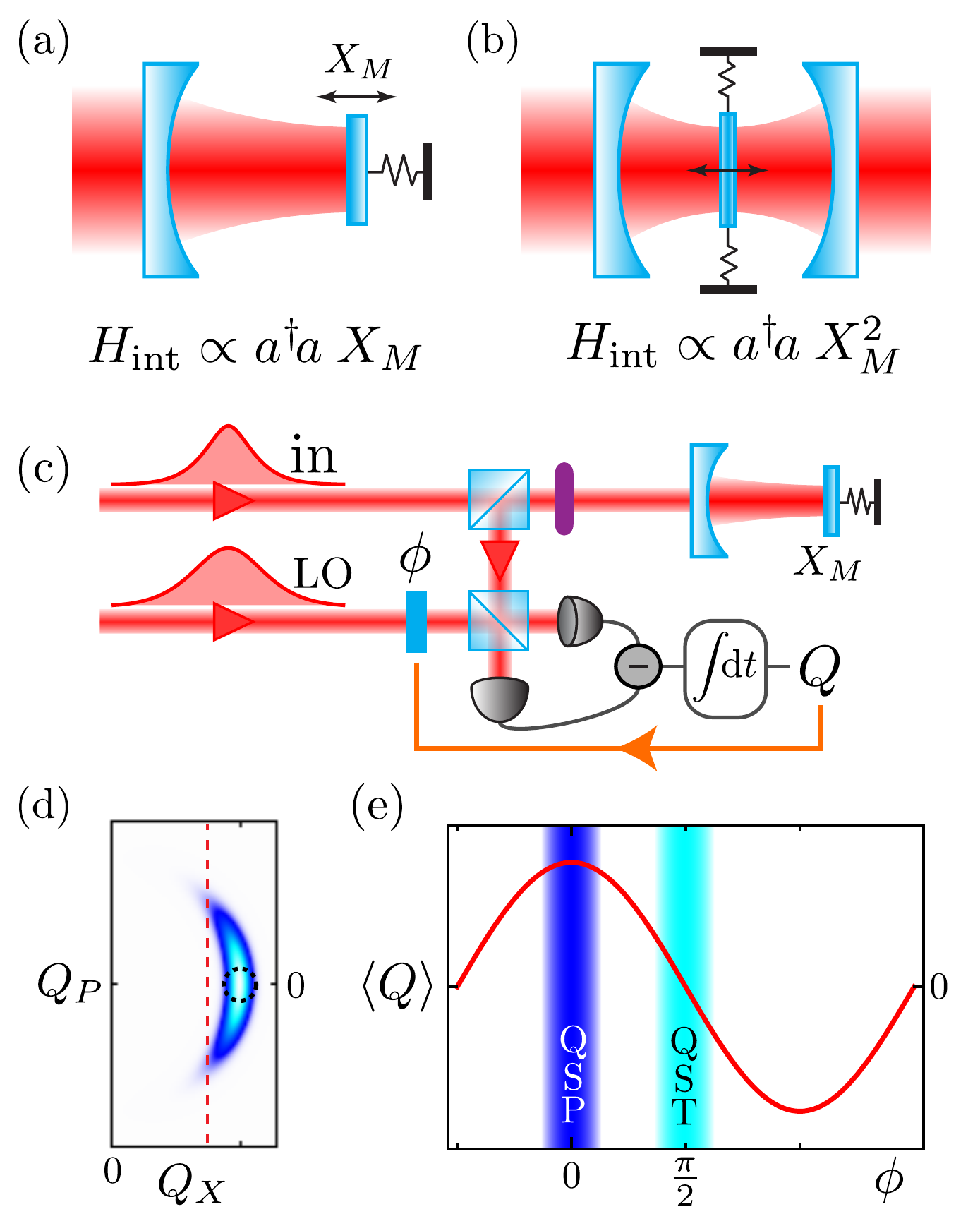}
\caption{
Cavity optomechanics is currently realized using `reflective' (a) and `dispersive' (b) approaches. The interaction in the former is proportional to mechanical displacement $\XM$, however, in the latter the interaction can be tuned to being proportional to $\XM$ or $\XM^2$. Pulses of light may be used to probe and manipulate the motional state of the mechanical resonator. Considered here is a pulse incident upon an optomechanical system with an interaction proportional to $\XM$ and then an optical quadrature measurement performed via homodyne detection (c). Following the interaction, the Wigner function of the optical field (d) is scimitar shaped due to mechanical position induced optical rotations. The distribution of the initial coherent state is indicated by the dashed circle. (The parameters for this plot were chosen to exaggerate the curvature.) For small rotations, the phase quadrature $\homP$ is proportional to $\XM$ and the amplitude quadrature carries $\XM^2$ information. The amplitude quadrature measurement outcome indicated by the red dashed line may have originated from two distinct mechanical positions, which provides a means for superposition preparation. By choosing the phase in homodyne detection (e) one can use the amplitude quadrature for $\XM^2$ measurements and quantum state preparation (QSP) or use the phase quadrature for $\XM$ measurements to perform quantum state tomography (QST).
}
\label{Fig:Scheme}
\end{figure}

Currently, the main approaches to cavity optomechanics \cite{ref:reviews} can be divided into two categories - `reflective' and `dispersive'. In each approach the mechanical and optical degrees of freedom are coupled via radiation pressure and the dependence of the cavity resonance frequency on the mechanical position. The first approach is depicted in Fig.~\ref{Fig:Scheme}(a), where the optical field is reflected from a mechanical element and the change in cavity frequency and hence interaction Hamiltonian are linearly proportional to the mechanical position. Optomechanical realizations of this approach include deformable Fabry-P\'{e}rot cavities and optical whispering gallery mode resonators, which are discussed in Ref.~\cite{ref:reviews}. The second approach is depicted in Fig.~\ref{Fig:Scheme}(b), here a mechanical element is positioned within an optical field and partial reflection from both sides gives rise to a dispersive interaction. In this arrangement the cavity frequency varies periodically with mechanical displacement. This can be used to give a linear or quadratic position dependent cavity frequency change if the mechanical element is positioned at an anti-node or node of the field respectively. The ability to select between linear or quadratic displacement coupling provides considerable versatility and thus `dispersive' optomechanics is an exciting candidate to observe and explore quantum mechanical phenomena of macroscopic resonators. Optomechanical realizations of this approach utilize a dielectric membrane \cite{Thompson2008} or trapped cold atoms \cite{Purdy2010}, positioned within an optical cavity and experimental work is underway to realize this with an optically trapped microsphere \cite{Li2011}. The quadratic mechanical position coupling offered by dispersive optomechanics provides a route to observe quantization in mechanical energy \cite{Thompson2008}. Moreover, such quadratic coupling can also be used for cooling and squeezing of the mechanical element \cite{Nunnenkamp2010} and it can be enhanced by using additional optical spatial modes, which also even allows quartic interaction \cite{Sankey2010}.

In this paper, a scheme is presented which uses the optomechanical interaction linearly proportional to the mechanical position where, despite the form of the interaction, effective couplings to displacement and displacement squared are achieved. Here optical pulses that are short compared to a mechanical period are used and the square displacement coupling is obtained by exploiting the non-linear optical dependence of the interaction. This interaction has been linearized in much of the present literature, but continuous non-linear optomechanical interaction has recently been studied resulting in non-classical states of light \cite{Rabl2011} and of the mechanical oscillator \cite{Nunnenkamp2011}. Also, in the spin ensemble community, working beyond the linear regime has been used for non-Gaussian quantum state preparation \cite{Massar2003}. The optomechanical setup considered here is shown in Fig. \ref{Fig:Scheme}(c), where an optical pulse in a coherent state interacts with an  optomechanical system and is then measured via homodyne detection. Following the interaction, Wigner reconstruction of the optical subsystem of the optomechanical entangled state, would yield a `scimitar state' shown in Fig. \ref{Fig:Scheme}(d). The form of this state can be readily understood as the mechanical position fluctuations (including quantum fluctuations) rotate the optical field. For small rotations, one sees from Fig. \ref{Fig:Scheme}(d) that measurement of the optical phase quadrature allows for a measurement of mechanical position, however, of particular interest here, measurement of the amplitude quadrature may give outcomes which could have resulted from two distinct mechanical positions. This is due to an effective displacement squared coupling, which can be used for non-Gaussian state preparation. In Ref. \cite{Vanner2010} it was discussed how measurement of the optical phase quadrature can be used to perform quantum state tomography of the motional state of the mechanical resonator and generate conditional mechanical squeezed states. Thus, being able to select between displacement and displacement squared measurements provides the tools to generate non-Gaussian mechanical quantum states and perform state reconstruction simply by choosing the phase in the homodyne interferometer as is shown in Fig. \ref{Fig:Scheme}(e).

\section{Model}

The optomechanical Hamiltonian with linear mechanical position coupling in the optical rotating frame at the cavity frequency including coherent resonant drive is
\begin{equation}
\frac{\Href}{\hbar} = \omegam \bd b - \gref \sqrt{2} \ad a \XM - i \sqrt{2\kappa N_p} \alphai(a-\ad),
\end{equation}
where the optomechanical coupling rate, which is realization dependent, is of the form $\gref = \omegaL \xzp / L$. The cavity field's resonance frequency, annihilation operator and amplitude decay rate are $\omegaL, a$ and $\kappa$ respectively, and $L$ is the cavity length. The mechanical zero-point extension is $\xzp = \sqrt{\hbar/2 m \omegam}$, where $\omegam, b, m$ and $\XM (\PM)$ are the mechanical element's eigenfrequency, annihilation operator, effective mass and position (momentum) quadrature, respectively, where a single mechanical mode is considered. The input pulse has mean photon number $N_p$ and is described by $\alphai$, the normalized envelope, i.e. $\integ{t}\alphai^2(t) = 1$, which is assumed real.

During the interaction, which is short with respect to a mechanical period, i.e. $\kappa \gg \omegam$, the mechanical position is considered constant and the optical and mechanical equations of motion can be solved independently of one another. Immediately after the pulse interaction the mechanical position is unchanged, i.e. $\XMout = \XMin$, however optomechanical entanglement is generated and correlations are established between the mechanical momentum and the optical intensity, $\PMout = \PMin + \sqrt{2}\gref \integ{t} \ad a$. 

The intra-cavity field evolves during the non-linear optomechanical interaction according to
\begin{equation}
\deriv{a}{t}  = ( i \gref \sqrt{2}\XM - \kappa ) a   + \sqrt{2\kappa} ( \sqrt{N_p}\alphai + \ai ),
\end{equation}
where the field is rotated in proportion to the mechanical position. This can be immediately solved exactly \cite{ref:MilburnNonLinear}, however, in this work the solution is approximated as the rotation is assumed small and the mean of the field is
\begin{equation}
\frac{\mean{a(t)}}{\sqrt{N_p}} \simeq  \alpha_0(t) + i \frac{\gref}{\kappa} \alpha_1(t) \mean{\XM} - \frac{\gref^2}{\kappa^2} \alpha_2(t) \mean{\XM^2},
\label{eq:meanOpticalField}
\end{equation}
where the dimensionless temporal mode functions $\alpha_{0,1,2}$ are introduced
\cite{ref:modeFunctions}. The phase quadrature of the intracavity field contains information on the mechanical displacement and the amplitude quadrature carries information of the mechanical displacement squared, Fig.~\ref{Fig:Scheme}(e). Measurement of these quadratures can be performed by time-domain homodyne detection of the output field $\ao = \sqrt{2\kappa}a - \ai$. Homodyning the amplitude quadrature is described by $\homX = \sqrt{2}\integ{t} \alphalo(t)\XLout(t)$, where $\XLout = 2^{-1/2}(\ao + \aod)$ (similarly $\homP$ describes phase quadrature detection). For an optimal measurement of $\XM^2$, ($\XM$) one chooses the local oscillator pulse $\alphalo$ to have an amplitude directly proportional to $\alpha_2$, ($\alpha_1$). The mean of the amplitude quadrature measurement is $\mean{\homX} = \homXZero - \chiX \langle\XM^2\rangle$, where the first term is the contribution from $\alpha_0$ and $\chiX$ is the square displacement measurement strength. For convenience, the homodyne measurement outcome is re-written as $\delhomX = \homXZero - \homX$. The optimal single pulsed measurement of $\XM$ is achieved with an input drive with a Lorentzian spectrum which matches the natural decay of the cavity \cite{Vanner2010}. The square displacement measurement strength is optimal when $\alphai^2(\omega) = (3\pi)^{-1} 8\kappa^5 / (\kappa^2 + \omega^2)^{3}$, which is not Lorentzian due to the higher order nature of the interaction considered here. This gives $\chiX = \sqrt{42 N_p} \gref^2 / \kappa^{2}$.

This kind of pulsed interaction and measurement is well suited to being described with the use of measurement operators as outcome probabilities and conditional mechanical states can be readily determined \cite{Wiseman2010}. Homodyne detection of the amplitude quadrature has the outcome probability density $\prob{\delhomX} = \traceSubInline{M}{\UpsdXL \UpsXL \rhoMin}$, where $\UpsXL$ is the corresponding measurement operator. In this pulsed regime $\delhomX$ has mechanical dependence only on $\XM$ which allows $\UpsdXL \UpsXL$ to be interpreted as an outcome probability density conditioned on a mechanical position. For the coherent optical drive considered here one obtains
\begin{multline}
\UpsXL(\XM, \delhomX) = \\
\pi^{-1/4} e^{i\OmegaRefX \XM} \exp\left[ -\frac{1}{2} (\delhomX - \chiX \XMval^2)^2 \right],
\end{multline}
where the mean momentum transfer is $\OmegaRefX = (5 \sqrt{2} / 3) N_p \gref / \kappa$.

\section{Comparison to the dispersive quadratic interaction}

Before proceeding to a discussion of the mechanical states of motion that can be prepared with $\UpsXL$, the square displacement measurement scheme introduced above is compared with the dispersive case. The Hamiltonian from Ref.~\cite{Thompson2008} for optomechanical systems with a dispersive element positioned so that the cavity frequency varies quadratically with the position of the element, in the optical rotating frame at resonance, including drive is
\begin{equation}
\frac{\Hdis}{\hbar} = \omegam \bd b + \gdis \ad a \XM^2 - i \sqrt{2\kappa N_p} \alphai(a-\ad),
\end{equation}
where the quadratic optomechanical coupling rate is $\gdis = (16 \pi^2 c \xzp^2 / L \lambda^2) \sqrt{2(1-r)}$, $r$ is the (field) reflectivity of the dispersive element and $\lambda$ is the optical wavelength. The phase quadrature of an optical pulse incident upon such an optomechanical system will be displaced in proportion to $\XM^2$ and it is readily shown that for a homodyne measurement of the phase quadrature with outcome $\homP$ the measurement operator is $\UpsDis = \pi^{-1/4} e^{- i \OmegaDis \XMin \XM} \exp[-\frac{1}{2}\left(\homP + \chiDis \XM^2 \right)^2]$, which has recently been used in Ref.~\cite{RomeroIsart2011}. After pulse shape optimization, $\OmegaDis = 3 N_p \gdis / \kappa$ and $\chiDis = \sqrt{10 N_p} \gdis / \kappa$. Comparing the measurement strengths for the dispersive direct $\XM^2$ interaction and the effective $\XM^2$ from the linear interaction for identical $N_p$ and $\lambda$ gives
\begin{equation}
\frac{\chiX}{\chiDis} \simeq \frac{1}{\pi} \frac{\Fref^2}{\Fdis} \frac{\xref^2}{\xdis^2} \frac{1}{\sqrt{2(1-r)}},
\end{equation}
where the cavity finesses and mechanical zero-point extensions are distinguished by subscripts for the two optomechanical cases. \emph{Remarkably, using the optomechanical interaction which is linearly proportional to $\XM$ and optical amplitude quadrature measurements gives an $\XM^2$ measurement strength which scales more favorably than that available with the direct $\XM^2$ interaction in dispersive optomechanics.} This, in combination with the measurement based selectability between linear or quadratic couplings offered here are the main results of this work.

\section{Experimental protocol and discussion}

Jacobs and colleagues discussed the preparation of superposition of the position of a mechanical resonator via $\XM^2$ measurements \cite{Jacobs2009}, work which has also recently been extended to include feedback control of the superposition separation \cite{Jacobs2011}. Such benchmark quantum states show striking differences between classical and quantum behavior and are thus highly sort experimentally to study the quantum mechanical properties of macroscopic objects \cite{Bose1999, Armour2002, Marshall2003}. In the following, an experimental protocol and a parameter set are discussed to prepare and observe spatial superposition of a massive mechanical resonator using the non-linear interaction and measurement $\UpsXL$. A measurement on a variety of experimentally accessible initial states is considered and the resulting conditional and unconditional mechanical states of motion are determined.

As the spectrum of measurement outcomes is continuous it is not experimentally possible to post-select from many experimental runs on a single outcome, rather a window must be used. The mechanical state conditioned on outcomes occurring in the window $\delhomX \pm w/2$ (labeled by $w$) is
\begin{equation}
\rhoM^{(w)} = \frac{1}{\prob{w}} \integlim{w}{\phantom{ }}{\delhomX'} \, \UpsXL(\delhomX') \rhoMin \UpsdXL(\delhomX'),
\label{eq:window}
\end{equation}
where $\prob{w}  = \integlim{w}{\phantom{ }}{\delhomX} \prob{\delhomX}$ is the probability of obtaining an outcome in the window. The mean measurement outcome for a mechanical thermal state with thermal occupation $\nbar$ is $\mean{\delhomX} = \chiX \left( 1 / 2 + \nbar \right)$. As the mean is greater than zero some insight is gained into the form of $\prob{\delhomX}$, which is a non-Gaussian function with a large wing for positive outcomes which increases for larger mechanical position variance. 

\begin{figure}[t]
\includegraphics[width=0.95\hsize]{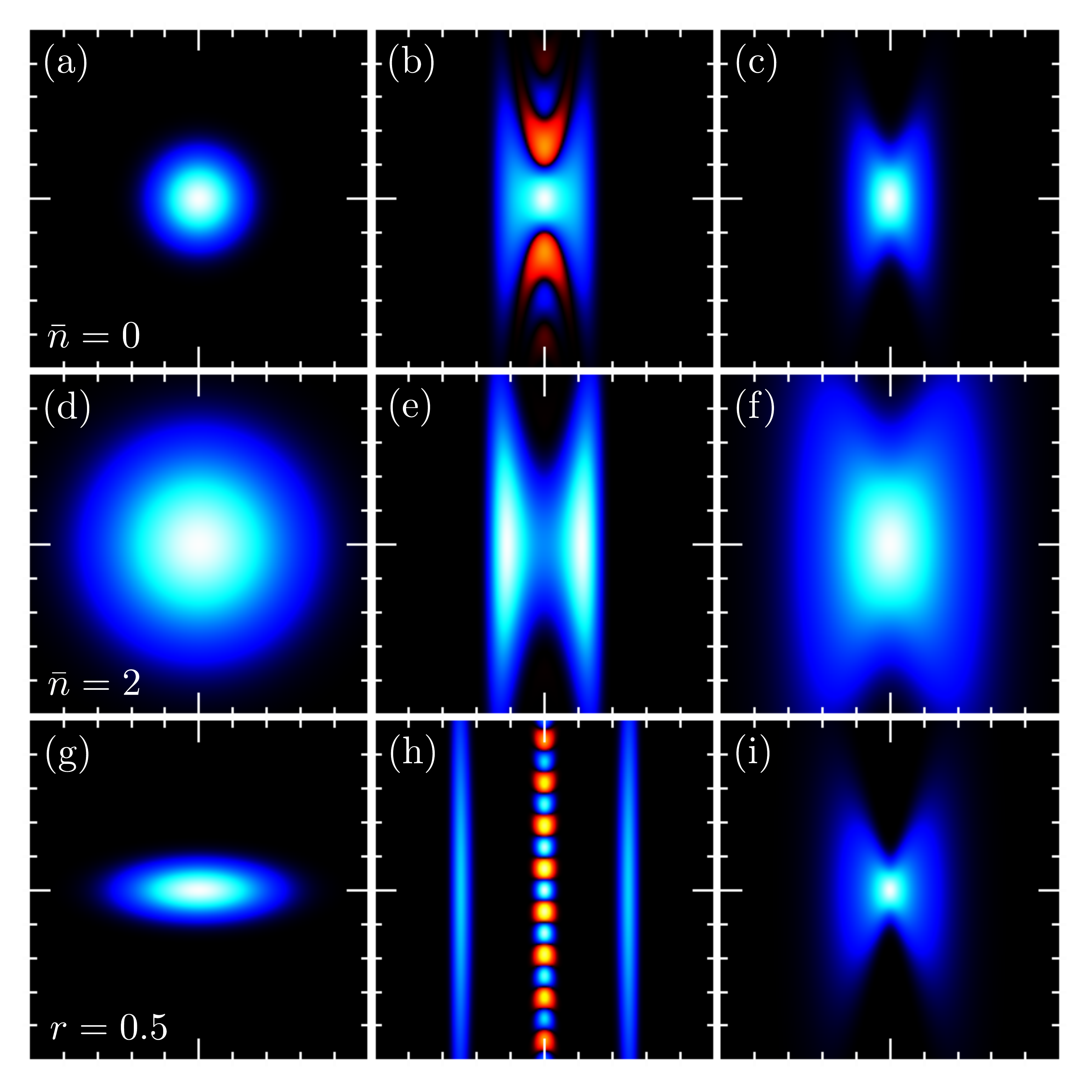}
\caption{
Mechanical Wigner functions of initial states (left), conditional states (center) and unconditional states (right). (The plot range is $\pm$5 for all axes. Color scale: black represents zero magnitude, blue for positive values and red for negative values.) The initial states are the ground-state, $\nbar = 0$ (a), a thermal state with $\nbar = 2$ (d) and a momentum squeezed vacuum state with squeezing parameter $r = 0.5$ (g). Conditional states prepared with $\Upsilon_{\! X}$ acting on the corresponding initial states with $\chiX = 1, \delhomX = 1.5, w = 0.8$ are shown in (b,e) and $\delhomX = 6.4$ has been used in (h). The probabilities of obtaining an outcome in the windows used above are: (b) 14.9~\%, (e) 14.5~\%, (h) 1.1~\%. Note the disappearance of negativity - a quantum to classical transition - for initial thermal occupation (e) and if the measurement outcomes are ignored (c,f,h).
}
\label{Fig:Wigner}
\end{figure}

In Fig.~\ref{Fig:Wigner} the action of $\UpsXL$ is considered on three mechanical Gaussian states: the ground-state, a thermal state and a momentum squeezed state. One may suspect that quite a narrow window for conditioning must be used in order to achieve significant coherence between the superposition components, however, conditional mechanical states, prepared from high-purity initial states, show strong quantum coherence even for relatively large conditioning windows. For example, the conditional mechanical state shown in Fig.~\ref{Fig:Wigner}(b) exhibits strong Wigner negativity even for $w = 0.8$, which allows the use of 15\% of the measurement outcomes. This plot also reveals the interesting feature that the negative regions are `curled around' positive regions, a feature which is not seen in the more commonly studied superposition of coherent states. This arises due to the population components having an asymmetric distribution about their peaks, specifically, there is a broader wing nearer to $\XMval = 0$ and a sharper edge on the other side. This form of the population components is more clearly seen in Fig.~\ref{Fig:Wigner}(e), which is the conditional mechanical state starting from a low occupation thermal state. When the population components have a more symmetric $\XMval$ distribution about their peak the interferences no longer curl as strongly, as is seen in Fig.~\ref{Fig:Wigner}(h) the conditional state starting from a squeezed state.

Since a measurement of the optical amplitude quadrature erases all the linear displacement information gained during the interaction and $\mean{\delhomX} > 0$, the unconditional (i.e. all measurement outcomes are ignored) mechanical state $\rhoMout = \integlim{-\infty}{\infty}{\delhomX} \UpsXL \rhoMin \UpsdXL$, is also non-Gaussian, however, mixed. 

The superposition separation $\delta$ is defined as the distance between the maxima of the two population components. This depends on the initial mechanical distribution, the measurement outcome and the square displacement measurement strength. For Gaussian initial mechanical states with a standard deviation $\sigma$ in their position spread, the superposition separation is
\begin{equation}
\delta = \frac{\sqrt{4 \delhomX \chiX - \sigma^{-2}}}{\chiX}.
\end{equation}
\begin{table}
\caption{An experimentally accessible set of parameters to achieve unity square displacement measurement strength.} 
\begin{tabular}{l c c c}
\hline
\hline 
Optical wavelength: & $\lambda$ & 1064 & [nm]\\
Mechanical effective mass: & $m$ & 40 & [ng]\\
Mechanical eigenfrequency: & $\omegam/2\pi$ & $2$ & [kHz]\\
Cavity finesse: & $\mathcal{F}$ & $5 \times 10^4$ & \\
Photon number per pulse: & $N_p$ & $1.7 \times 10^{9}$ & \\
\hline 
Cavity length: & $L$ & $750$ & [$\mu$m]\\
Mechanical ground-state size: & $\xzp$ & $10$ & [fm]\\ 
Optomechanical coupling: & $\gref/2\pi$ & $3.8$ & [kHz]\\
Single photon strength: & $\gref/\kappa$ & $1.9 \times 10^{-3}$ & \\
Quadratic pos. meas. strength: & $\chiX$ & $1.0$ & \\
Separation ($\nbar=0, \delhomX=1.5$): & $\delta$ & 2.0 & \\
\hline
\hline
\end{tabular}
\label{Tab:Parameters}
\end{table}

Experimental progress in optomechanics is steadily approaching the regime where the important parameter $\gref / \kappa$, which quantifies the mechanical momentum displacement by a single photon (for $\kappa \gg \omegam$) approaches unity. In this work $\chiX$ scales with the square of this parameter and the pulsed position measurement strength for mechanical quantum state tomography~\cite{Vanner2010} scales linearly with this parameter. In present-day experiments however~\cite{ref:ExpsGoK}, $\gref / \kappa \ll 1$, which this work overcomes by utilizing large coherent amplitudes in order to achieve sufficient coupling to prepare and observe non-Gaussian mechanical states of motion. To ensure short interaction the cavity decay rate is chosen as  $\kappa = 10^3~\omegam$, which for a desired finesse sets the cavity length required. In Table~\ref{Tab:Parameters} a list of parameters is provided for a deformable Fabry-P\'{e}rot optomechanical system with a kHz scale mechanical resonator.

The protocol for quantum state preparation and quantum state tomography comprises three steps: \emph{i}) an initialization stage of mechanical pre-cooling and/or squeezing. Since $\kappa \gg \omegam$ is required here and low frequency mechanical resonators are considered active-feedback cooling is most suitable \cite{Genes2008, Zhang2009}. Alternatively, in this regime squeezing and purification can be achieved with the use of conditional measurement~\cite{Vanner2010}. Additionally, squeezing can be achieved by applying a parametric modulation to the mechanical device \cite{Mari2009}. \emph{ii}) Following this, a pulse is injected into the optomechanical cavity to realize $\UpsXL$ and the measurement outcome is recorded. At this point, the mechanical oscillator has gained the momentum $\OmegaRefX$ which after one quarter of a period of free evolution shifts the cavity resonance frequency by $\Delta \omega_\Omega = \gref \sqrt{2} \OmegaRefX$. As this can be much larger than $\kappa$ any subsequent pulse will not resonantly drive the cavity. In order to overcome this, a two pulse preparation sequence can be used where a second pulse follows after half a mechanical period of free evolution to cancel the mean momentum gained by the resonator. In this case one applies $\UpsXL$ twice where both outcomes are recorded, thus strengthening the measurement of $\XM^2$. This procedure requires a good degree of optical amplitude stability, which is anyway necessary for $\UpsXL$ measurements. During the free evolution the appropriate master equation is solved to determine the mechanical state immediately prior to the second measurement. However, as discussed below, given the parameters considered here the mechanical bath coupling is not expected to play a strong role during this timescale. \emph{iii}) With the resonator state near the origin of phase-space, quantum state tomography, as discussed in Ref.~\cite{Vanner2010}, is now performed. This is achieved here by later injecting a subsequent pulse with the local oscillator phase switched to measure the optical phase quadrature as in Fig.~\ref{Fig:Scheme}(e). Repeating this protocol many times and post-selecting the measurement outcomes $\delhomX$ within the desired window provides a powerful experimental platform to generate and fully reconstruct a non-Gaussian state of motion of a mechanical resonator.

In order to prepare mechanical superposition states with $\UpsXL$ there needs to be sufficient mechanical displacement induced optical rotation such that two distinct positions give the same amplitude quadrature outcome. This is best achieved if the mechanical mean position gives zero rotation. For mechanical states that have a non-zero mean, which could have been conditionally prepared with a prior pulse \cite{Vanner2010}, non-Gaussian state preparation and tomography can be performed by providing a feedback phase-shift (indicated by the arrow in Fig.~\ref{Fig:Scheme}(c)) to rotate the optical scimitar to be centered about the $\homX$ axis, as in Fig.~\ref{Fig:Scheme}(d). Additionally, it is remarked that for optical rotation beyond that considered in (\ref{eq:meanOpticalField}) existing experimental calibration procedures and the interpretation of optical phase measurements will require modification to take the optomechanical non-linearity into account.


Studying the decoherence of quantum superposition in a mechanical resonator is an important step to determine the feasibility of optomechanical systems as components for quantum information applications. Proposals for such applications are numerous and include: quantum memory~\cite{Zhang2003}, optomechanically mediated qubit-light transduction~\cite{Stannigel2010} and coherent optical wavelength conversion~\cite{Tian2010}, to name a few. There is much literature on the topic of environmental coupling and decoherence~\cite{ref:decoherence} and so no detailed discussion will be provided here. However, in the context of this proposal, what is important is the parameter $\nbar/Q$, where $Q$ is the mechanical quality factor. This parameter quantifies the rate of rethermalization normalized to the mechanical frequency and must be much less than unity to study the evolution of quantum mechanical phenomena over the timescale $\omegam^{-1}$. For a temperature of 25~mK accessible with dilution refrigeration and a $Q=5\times 10^6$ this gives $\nbar/Q = 0.05$ using the mechanical frequency above. With the full quantum state tomography available here this scheme allows the dynamics of mechanical superposition states to be measured, which may be used to characterize the couplings responsible for decoherence thus allowing for improved mechanical device engineering.

Furthermore, the significant mass involved in the spatial superposition offers a parameter regime that allows for an experimental test of collapse models. Very recent proposals in matter-wave interferometry~\cite{ref:matterWave, RomeroIsart2011}, which also consider the use of filtering type measurements to generate superposition, may provide the ability to test continuous spontaneous localization (CSL)~\cite{ref:CSL}. The mechanical resonator parameters considered here are not suitable to test CSL, predominantly as the superposition separation is small~\cite{Bassi2005}. However, the separation can be larger than the distribution of the mass contained within the nucleus and so this can be used to test gravitational collapse~\cite{ref:GravCollapse}. For example, using the parameters above ($\delta = 2.0, \xzp = 10$~fm) the separation is approximately 28~fm and the diameter of a $^{28}$Si nucleus is approximately 8~fm. It may be useful in such an investigation to start with an initial squeezed mechanical state, as is considered in Fig.~\ref{Fig:Wigner}(g-i), as one can study a larger range of superposition separations as the probability density of measurement outcomes is broader.

\section{Conclusion}

This work has provided a means to measure displacement or displacement squared of a mechanical resonator using the optomechanical interaction linearly proportional to mechanical displacement by simply changing the phase in optical homodyne measurement. Displacement squared measurements have so far been predominantly considered in dispersive optomechanics, however, the optimal square displacement measurement strength obtained in the scheme introduced here scales more favorably than that available in dispersive optomechanics. This opens the possibility that  optomechanics with a linear mechanical coupling may also provide a route to observe mechanical energy quantization, as was considered in \cite{Thompson2008}. Moreover, as was proposed in \cite{Jacobs2009}, with an $\XM^2$ coupling to a mechanical resonator one can prepare a superposition of positions via measurement. This, applied to the $\XM^2$ coupling achieved here and combined with the ability to perform mechanical state tomography with time \cite{Vanner2010} provides an alternative to Refs. \cite{Bose1999} and \cite{Marshall2003} to generate superposition of a mechanical resonator without the need for large single photon mechanical displacement $\gref/\kappa$. Such mechanical superposition states are important to experimentally investigate in order to determine the feasibility of mechanical resonators as elements in quantum information applications and to explore decoherence mechanisms arising from environment interaction or, for example, gravitationally induced collapse.

\section{Acknowledgments}
M.R.V. is a member of the FWF Doctoral Programme CoQuS (W 1210), is a recipient of a DOC fellowship of the Austrian Academy of Sciences and gratefully acknowledges discussion with Markus Aspelmeyer, Gerard J. Milburn and Igor Pikovski.


\end{document}